%% file: paper.tex
\documentclass[aps,prl,showpacs,amsmath,amssymb,amsfonts,lengthcheck,twocolumn,longbibliography,superscriptaddress]{revtex4-2}

\usepackage[T1]{fontenc}
\usepackage{graphicx}
\usepackage{physics}
\usepackage{amsthm}
\usepackage{amsmath}
\usepackage{verbatim}
\usepackage{dcolumn}
\usepackage{bm}
\usepackage{epsf}

\usepackage{color}
\usepackage[colorlinks=true,citecolor=blue,linkcolor=blue,urlcolor=blue]{hyperref}%
\usepackage{xcolor}
\usepackage{dsfont}
\usepackage{tikz}
\usepackage{pgfplots}
\usepackage{xcolor}
\usepackage{url}
\usepackage{placeins}
\usepackage[normalem]{ulem} 

\definecolor{myvermillion}{HTML}{D55E00}
\definecolor{myyellow}{HTML}{F0E442}
\definecolor{mygreen}{HTML}{009E73}
\definecolor{mypurple}{HTML}{CC79A7}
\definecolor{myblue}{HTML}{0072B2}
\definecolor{myorange}{HTML}{E69F00}
\definecolor{mygray}{RGB}{153,153,153}

\pgfplotsset{compat=newest}

\newcommand{\ptr}[2]{\mathrm{tr_{#1}}\left\{#2\right\}}

\newcommand{\bla}{bla\\bla\\bla\\bla\\bla}

\newcommand{\mrm}[1]{\mathrm{#1}}

\makeatletter
\newcommand{\currentfontsize}{\f@size pt}
\makeatother

\usepackage[colorlinks=true,citecolor=blue,linkcolor=blue,urlcolor=blue]{hyperref}

\begin{document}

\title{Quantum timekeeping and the dynamics of scrambling in critical systems}

\author{Devjyoti Tripathy}
\email{dtripathy@umbc.edu}
    \affiliation{Department of Physics, University of Maryland, Baltimore County, Baltimore, MD 21250, USA}
    \affiliation{Quantum Science Institute, University of Maryland, Baltimore County, Baltimore, MD 21250, USA}

\author{Federico Centrone}
\affiliation{ICFO-Institut de Ciencies Fotoniques, The Barcelona Institute of Science and Technology, Av. Carl Friedrich Gauss 3, 08860 Castelldefels (Barcelona), Spain}
\affiliation{Universidad de Buenos Aires, Instituto de Física de Buenos Aires (IFIBA), CONICET,
Ciudad Universitaria, 1428 Buenos Aires, Argentina}

\author{Sebastian Deffner}
    
    \affiliation{Department of Physics, University of Maryland, Baltimore County, Baltimore, MD 21250, USA}
    \affiliation{Quantum Science Institute, University of Maryland, Baltimore County, Baltimore, MD 21250, USA}
    \affiliation{National Quantum Laboratory, College Park, MD 20740, USA}
\date{\today}

\begin{abstract}
    In this work, we develop a quantum metrological framework for quantum chaos by showing that local subsystems of information scrambling systems naturally function as quantum stopwatches. The reduced quantum state of a subsystem encodes the passage of time through its growing distinguishability from the initial preparation. Treating time as the estimation parameter, we then derive a generalized quantum Cramer-Rao bound that directly relates the precision of time estimation to the decay of out-of-time ordered correlators (OTOCs) and subsystem quantum Fisher information (QFI). As a main result, we obtain continuity bounds for quantum Lyapunov exponent in terms of the subsystem QFI in quantumly chaotic dynamics. Furthermore, using a scaling analysis based on imaginary-time correlators, we show that the subsystem QFI exhibits universal critical amplification near quantum phase transitions. Our results are demonstrated and verified by a numerical analysis of the dynamics of a chaotic Ising chain.
\end{abstract}

\maketitle


Without any doubt, timekeeping is one of the most fundamental tasks in everyday live as well as in science and technology \cite{de-timekeeping}. From tracking planetary motion \cite{Bahinipati-sundial-temple} and enabling GPS \cite{cao-gps-satellite} to probing frequency changes in atomic transitions \cite{optica-atomicclocks-review,zhang-atomic-clock-nature} and understanding the invisible universe \cite{safronova-astro-time,safronova-darkmatter-time}, time-keeping is not merely about reading clocks, it is about extracting temporal information with maximal precision \cite{woods-qclocks-prx}. At the quantum level, timekeeping acquires an even deeper meaning: the ability of a quantum system to function as a timekeeping device, i.e. stopwatch, is determined by how sensitively its state evolves according to the quantum dynamics. In quantum metrology, this sensitivity is quantified by the quantum Fisher information (QFI) \cite{Liu-qfireview-iop, helstrom-1969-qestimation, care-quantum_stats-book} that sets the ultimate precision with which a parameter, such as time, can be estimated from measurements via the quantum Cramer-Rao bound \cite{yu-qcr-verify-npj,Liu-qfireview-iop,albarelli-qcr-prl,toth-qcr_review-iop}.

The obvious question arises with what quantum systems the most accurate time-keeping devices can be built. This question is far from trivial, since in complex many-body systems quantum dynamics generically leads to the spread of quantum information and growth of entanglement \cite{Islam-ent_entropy-nature,Zhang-ent_coherence-npj,Schweigler-high_corr-nature}. Even if the global dynamics is unitary, local degrees of freedom rapidly loose their memory of initial conditions, giving rise to effectively irreversible behavior in smaller subsystems \cite{dev-qsl_otoc-pra}. This so-called \emph{quantum information scrambling} lies at the heart of quantum chaos, thermalization, and non-equilibrium dynamics \cite{Shenker-blackholes_butterfly-jhep,hayden-black_holes-jhep,goldfriend-quasi_therm-pre,xu-chaos_scramble-prl,swingle-scrambling-prx,Hallam-lyapunov_thermal-nature,Touil2024EPL}. 

A widely used diagnostic of information scrambling is the out-of-time ordered correlator (OTOC), a four-point correlator that quantifies how strongly two initially local operators fail to commute under time evolution \cite{Anand-brotocs-quantum,Dong-scarred-PRR,Swingle-otocs-NP,cotler-otoc_butterfly}. Physically, the OTOC measures the sensitivity of late-time observables to early-time perturbations, and thus provides a quantum analog of classical chaos \cite{akutagawa2020out}. For quantum chaotic systems \cite{dev-otoc_convul-aip}, the OTOC exhibits a simple exponential decay, $\text{OTOC}(t)\sim \exp{-\lambda_\mrm{Q} \,t}$, where $\lambda_\mrm{Q}$ is called the quantum Lyapunov exponent \cite{Maldacena-bound_chaos-jhep,cameo-lyapunov-pre,Hallam-lyapunov_thermal-nature,garcia-lyapunov-prd}.

These two perspectives, metrological sensitivity and information scrambling, are usually studied independently \cite{braun-chaos_metrology-nature}. Yet, they raise a natural and intriguing question: if QFI quantifies how accurately a quantum system can tell time  and scrambling describes how rapidly quantum information spreads through a many-body system, can the most precise quantum stopwatches realized in systems that scramble information the most efficiently?

In the present letter, we answer this question affirmatively. Rather than treating scrambling purely as operator growth, we adopt the perspective of a local observer. To this end, we consider a subsystem $\mathcal{A}$ embedded in a larger interacting quantum many-body system $\mathcal{S}$. Although the global dynamics is unitary, the reduced state of $\mathcal{A}$ evolves in a highly non-trivial manner due to its coupling with the rest of the system $\mathcal{S}$. From the perspective of the subsystem $\mathcal{A}$, the passage of time is encoded in the growing distinguishability of its reduced, time-evolved state from its initial preparation. The subsystem, therefore, acts as a \emph{natural quantum stopwatch} whose precision is determined by its time-dependent QFI.
Building on this perspective, we derive a general quantum Cramer-Rao bound that connects metrological sensitivity to the dynamics of scrambling. For a single qubit subsystem, this relation yields a bound on the quantum Lyapunov exponent in terms of a time-averaged subsystem QFI. Therefore, our result establishes a direct link between quantum chaos and metrology: the faster information scrambles in a many-body system, the more precisely its local subsystems can function as quantum stopwatches.

Our framework also provides a theoretical explanation for the previously conjectured peak of the Lyapunov exponent near quantum critical points \cite{shen-lyapunov_peak-prb}. By relating the subsystem QFI to imaginary-time correlation functions, we show that the QFI exhibits universal amplification in the vicinity of a quantum phase transition. This amplification reflects the enhanced temporal distinguishability induced by critical fluctuations and provides a concrete physical mechanism for the conjectured peak of the quantum Lyapunov exponent near criticality. 

Together, these results establish a conceptual bridge between three central concepts: quantum chaos, criticality and metrology, identifying them as different manifestations of the same underlying dynamical structure.

\paragraph{Generalized Quantum Cramer-Rao Bound}

We consider a quantum many-body system $\mathcal{S}$ evolving under unitary dynamics induced by the Hamiltonian $H$. Further, $\mathcal{S}$ comprises two subsystems $\mathcal{A}$ and $\mathcal{B}$, of which $\mathcal{A}$ plays the role of ``subsystem of interest'', and $\mathcal{B}$ becomes the ``effective environment''.

To construct a quantum stopwatch out of $\mathcal{A}$, we now need to consider an observable $A$, that lives on $\mathcal{A}$, but not on $\mathcal{B}$. For such observables, a quantum speed limit can be written as \cite{garcia-unifying_qsl-prx}
\begin{equation}
\label{eq:garcia_qsl}
    \frac{d}{dt}\langle A\rangle \leq \Delta A\,\sqrt{I_{F}^{\mathcal{A}}(t)}
\end{equation}
where the $I_{F}^{\mathcal{A}}(t)$ is the subsystem QFI of $\mathcal{A}$ with time as the estimation parameter and $(\Delta A)^2 = \langle A^{2}\rangle - \langle A\rangle^{2}$ is the variance of observable $A$. Note, that all expectation values are taken with respect to $\mathcal{A}$. 

For the reduced density matrix of $\mathcal{A}$, $\rho_{\mathcal{A}}(t) = \sum_j p_j\ket{j}\bra{j}$, the QFI $I_{F}^{\mathcal{A}}(t)$ reads \cite{tang-time_estimation_scipost}
\begin{equation}
\label{eq:qfi}
\begin{split}
    I_{F}^{\mathcal{A}}(t) &=2\sum_{jk}^{d}\frac{\abs{\bra{j}\dot{\rho}_{\mathcal{A}}\ket{k}}^{2}}{p_j+p_k}\\
    &= 2\sum_{jk}^{d}\frac{\abs{\bra{j}\ptr{\mathcal{B}}{[H,\rho(t)]}\ket{k}}^{2}}{p_j+p_k}
\end{split}
\end{equation}
where the dot denotes a derivative with respect to time.

To continue, we now need to judiciously choose the observable $A$. As we will see shortly and motivated by standard treatments of the quantum speed limit \cite{Deffner2017JPA}, we choose $A(t)=\rho_{\mathcal{A}}(t)$. In this case, Eq.~\eqref{eq:garcia_qsl} becomes
\begin{equation}
\label{eq:obs_den}
    \frac{d}{dt}\tr{\rho_{\mathcal{A}}^2(t)}\leq 2\Delta\rho_{\mathcal{A}}(t)\sqrt{I_{F}^{\mathcal{A}}(t)}
\end{equation}
where $\left(\Delta\rho_{\mathcal{A}}\right)^2 = \tr{\rho_{\mathcal{A}}^3}-\tr{\rho_A^2}^2$. For ease of notation, we drop the superscript $\mathcal{A}$ from the subsystem QFI and refer to it as $I_{F}(t)$ in the rest of this analysis.

As we will see shortly, Eq.~\eqref{eq:obs_den} can be written in particularly appealing from in the context of quantum information scrambling. To this end, recall the definition of the OTOC,
\begin{equation}
\label{eq:otoc4pt}
    \text{OTOC}(t)=\langle B^{\dagger}(t)A^{\dagger}B(t)A\rangle_{\rho},
\end{equation}
where $A$ and $B$ are local operators acting on subsystems $\mathcal{A}$ and $\mathcal{B}$, and $\rho$ is the initial state of the composite system $\mathcal{A}\otimes\mathcal{B}$. The operator $B(t)=\exp{iHt}\,B\,\exp{-iHt}$ represents the time-evolution of $B$ in the Heisenberg picture under the Hamiltonian of the total system $H$, and we set $\hbar=1$.

To connect metrological sensitivity to information scrambling, we now use the fact that the averaged OTOC can be expressed in terms of the Renyi-2 entropy of a subsystem \cite{fan-otoc_re-scibull,zhou-otoc_re-arxiv}. In fact, the averaged OTOC on the composite system is related to purity of subsystem $\mathcal{A}$ simply by
\begin{equation}
    \label{eq:otoc-re}
    \mathcal{O}_t = \exp{-S_\mathcal{A}^{(2)}(t)}\,,
\end{equation}
where
\begin{equation}
 S_\mathcal{A}^{(2)}(t) = -\ln\left(\tr\left\{\rho_\mathcal{A}^2(t)\right\}\right)
\end{equation} 
is the Renyi-2 entropy of subsystem $\mathcal{A}$. The averaged OTOC, $\mathcal{O}_t$ is defined as a Haar average over all unitaries on subsystem $\mathcal{B}$
\begin{equation}\label{eq:avg-otoc}
    \mathcal{O}_t = \int dB \, \tr\left\{B^{\dagger}(t)A^{\dagger}B(t)A\right\},
\end{equation}
where $dB$ denotes the Haar measure. 

Combining Eqs.~\eqref{eq:otoc-re} and \eqref{eq:obs_den}, we obtain
\begin{equation}\label{eq:gqcr}
    (\Delta\rho_A)^2 \geq \frac{\dot{\mathcal{O}}_t^{2}}{4I_{F}(t)},
\end{equation}
which can be understood as a generalized quantum Cramer-Rao bound.

In quantum metrology, $I_{F}^{\mathcal{A}}(t)$ sets the ultimate precision with which a parameter can be estimated. For any unbiased estimator $f(t)$, the standard quantum Cramer-Rao bound, reads
\begin{equation}
\label{eq:qcrb}
    (\Delta f(t))^2\geq \frac{1}{NI_{F}^{\mathcal{A}}(t)}
\end{equation}
where $I_{F}^{\mathcal{A}}(t)$ measures the growing distinguishability of an evolving quantum state.  This standard quantum Cramer-Rao bound \eqref{eq:qcrb} is purely metrological: it places a limit on estimation precision. By contrast, our  generalized quantum Cramer-Rao bound \eqref{eq:gqcr} plays a fundamentally different role. It captures how the metrological sensitivity of the subsystem limits the rate with which information can scramble within the global many-body system.

\paragraph{Bound on the quantum Lyapunov exponent} 

The generalized Cramer-Rao bound \eqref{eq:qcrb} holds for any quantum dynamics. However, the inequality becomes particularly relevant in quantumly chaotic systems, for which it can be leveraged to bound the Lyapunov exponent. To this end, consider now a specific subsystem $\mathcal{A}$ that consists of only a single qubit.

As we show in the Supplemental Material \cite{sm}, the variance of $\rho_{\mathcal{A}}(t)$ can then be expressed directly in terms of subsystem purity $S_{\mathcal{A}}^{(2)}(t)$, namely we have
\begin{equation}
\label{eq:var}
    \left(\Delta\rho_{\mathcal{A}}\right)^2 = \frac{1}{2}\left(3\,\mathcal{O}_t - 1\right) - \mathcal{O}_t^2
\end{equation}

Plugging Eq.~\eqref{eq:var} back into the generalized quantum Cramer-Rao bound Eq.~\eqref{eq:gqcr} and rearranging terms, we can write
\begin{equation}
\label{eq:qfi-otoc}
\quad
I_F(t)\;\ge\;\frac{ \dot{\mathcal{O}}_t^2}{2(3\mathcal{O}_t - 1 - 2 \mathcal{O}_t^2)}\,,
\quad
\end{equation}
which constitutes a differential inequality. The latter can be solved, and we finally obtain
\begin{equation}
\label{eq:diff-ineq}
    \mathcal{O}_t\leq \frac{1}{4}\cos{\left(2\int_0^t ds\,\sqrt{I_{F}(s)}\right)} + \frac{3}{4}\,.
\end{equation}
Note that Eq.~\eqref{eq:diff-ineq} is exact, and holds for any type of quantum dynamics of single qubit systems.

In chaotic scenarios, the OTOC decays exponentially \cite{Maldacena-bound_chaos-jhep,cameo-lyapunov-pre,Hallam-lyapunov_thermal-nature,garcia-lyapunov-prd}, i.e. $\mathcal{O}_t = \exp{-\lambda_Q t}$. In this case, and expanding for small times $t$, we have in leading order
\begin{equation}
\label{eq:final_lyapunov_bound}
\lambda_Q  \ge \frac{1}{2t}\left(\int_0^t ds\,\sqrt{I_F(s)}\right)^2\,.
\end{equation}
It is interesting to note that the right hand side of Eq.~\eqref{eq:final_lyapunov_bound} is nothing but the quantum speed limit (QSL) as formulated in Ref.~\cite{Taddei2013PRL}. In fact, we can also write
\begin{equation}
\lambda_Q  \ge \frac{2v_\mrm{QSL}^2}{ t}\,,
\end{equation}
where $v_\mrm{QSL}\equiv \int_0^t ds\,\sqrt{I_F(s)}/2$. In earlier work \cite{dev-qsl_otoc-pra}, some of us found an upper bound on the Lyapunov exponent in terms of the QSL. Here, we have now obtained a \emph{lower} bound. Thus, in quantumly chaotic dynamics the QSL provides continuity bounds on $\lambda_Q$.

Moreover, the present result \eqref{eq:final_lyapunov_bound} also has a very appealing interpretation. In fact, the QSL is the maximal geometric velocity of $\rho_{\mathcal{A}}(t)$ on the statistical manifold of density operators \cite{spehner-buresgeodesics}. Physically, this means that the rate at which information can scramble through a quantum many-body system is fundamentally controlled by the geometric speed and hence metrological distinguishability of local reduced states. In turn, the rate which which quantum states become distinguishable is the highest in quantumly chaotic dynamics. Thus, we immediately conclude that quantum stopwatches with the highest precision ought to be made from chaotic quantum systems.

\begin{figure}
\centering
\includegraphics{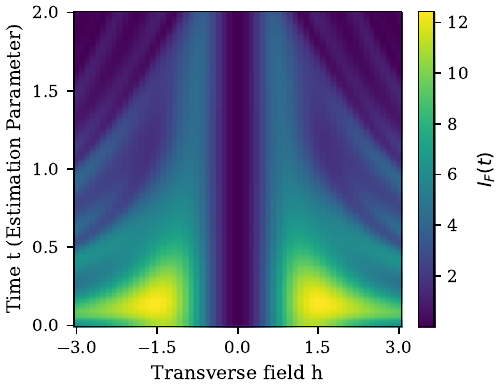}
\caption{Heatmap of $I_{F}(t,h)$ for the one-dimensional TFIM with a longitudinal field with open boundary conditions. $I_{F}(t)$ peaks approximately around $h=\pm J$ and decays over time. The parameters are $J=1.0$, $g=0.4$, $N=11$.}
\label{fig:subsystem qfi heatmap}
\end{figure}

\paragraph{Lyapunov exponent around criticality}

Now that we have determined that quantumly chaotic systems could provide the most accurate stopwatches, the natural question arise if there is a further distinction among chaotic systems To this end, recognize that any physical mechanism that enhances $I_{F}(t)$ will also directly enhance the achievable Lyapunov exponent. 

In the remainder of this analysis, we will show that quantum criticality provides precisely such a mechanism. Near a quantum phase transition, many-body systems exhibit long range correlations and enhanced fluctuations across all length scales \cite{sachdev-qpt-book}. These critical fluctuations strongly affect the reduced dynamics of local subsystems, which in turn modifies the time distinguishability of their reduced states. Since $I_{F}(t)$ quantifies this temporal distinguishability, we would expect the QFI to display characteristic scaling behavior near a critical point \cite{steve-metrology_qpt-pra}.

To analyze this behavior, we now relate $I_{F}(t)$ to the operator $M_{\mathcal{A}}(t) = \ptr{\mathcal{B}}{[H,\rho(t)]}$ which governs the effective non-unitary dynamics of the subsystem induced by its coupling to the remainder of the system.

In the Supplemental Material \cite{sm} we prove rigorously that we have
\begin{equation}
\label{eq:qfi-sandwich_rewrite}
    \frac{1}{2p_{\max}}\norm{M_{\mathcal A}(t)}_2^2\ \le\ I_F(t)\ \le\ \frac{1}{2p_{\min}}\norm{M_{\mathcal A}(t)}_2^2,
\end{equation}
where $p_{\max}$ ($p_{\min}$) denotes the maximum (minimum) eigenvalue of $\rho_{\mathcal A}(t)$.
Thus, we have shown that the $I_{F}(t)$ is upper and lower bounded by the Hilbert-Schmidt norm of the generator of the effective non-unitary dynamics. This is particularly useful, since it permits our problem to be recast in terms of imaginary-time two-point correlation functions of operators that couple subsystem $\mathcal{A}$ with subsystem $\mathcal{B}$. 

Near a quantum critical point, such correlators exhibit universal scaling behavior \cite{zanardi-critical_scaling-prl}. As also discussed in the Supplemental Material \cite{sm}, performing a standard scaling analysis shows that the intensive quantity
\begin{equation}
    m_{\mathcal A}^2(t)=L^{-d}\|M_{\mathcal A}(t)\|_2^2
\end{equation}
scales as
\begin{equation}\label{eq:scaling_qfi}
    m_{\mathcal A}^2(t)\sim\ |\lambda-\lambda_c|^{\Delta_M/\Delta_\lambda},
\end{equation}
in the vicinity of the critical point, where $\lambda$ is the tuning parameter and $\lambda_c$ denotes its critical value. For  the correlation length $\xi$  we further have $\xi = \abs{\lambda-\lambda_c}^{-\nu}$, where $\nu$ is the correlation length critical exponent. Then, with $\Delta_{\lambda}$ as the scaling dimensions of the tuning parameter $\lambda$, we can write $\nu = \Delta_{\lambda}^{-1}$. Combining Eqs.~\eqref{eq:qfi-sandwich_rewrite} and \eqref{eq:scaling_qfi}, we conclude that the $I_{F}$ exhibits quantum critical amplification under the condition $\Delta_M<0$.

Finally, inserting the resulting growth of $I_F$ into our bound Eq.~\eqref{eq:final_lyapunov_bound}, we unveil a concrete physical mechanism for the observed peak of the quantum Lyapunov exponent near quantum critical points \cite{shen-lyapunov_peak-prb}. Critical fluctuations amplify the local temporal distinguishability, which in turn enhances $\lambda_Q$ to its peak near its vicinity.


\paragraph{Numerical Example}

We conclude the discussion with a pedagogically valuable example. Namely, we demonstrate the validity of our results for the one-dimensional transverse field Ising Hamiltonian with a longitudinal field and open boundary conditions \cite{sharma-qdpt-prb, peng-scars_ising-prb}. The corresponding Hamiltonian reads,
\begin{equation}
\label{eq:chaotic_tfim}
    H = -J\sum_{i=1}^{N-1}\sigma_{i}^z\sigma_{i+1}^z - h\sum_{i=1}^{N} \sigma_{i}^x - g\sum_{i=1}^{N} \sigma_{i}^{z}\,.
\end{equation}
This chain has $N$ spin-$1/2$ nearest-neighbor interacting particles with Pauli matrices $\sigma_{i}^{\alpha}$, where $\alpha = x,y,z$ is the $\alpha$th component at site $i$. In the absence of the longitudinal field $g=0$, we have the usual quantum Ising chain with a transverse field which is integrable and can be mapped to a free fermion model via the Jordan-Wigner transformation and has a quantum critical point at $h_c=\pm J$ \cite{pierre-ising_integrable-ap,sachdev-qpt-book}. 

In the presence of the longitudinal field $g\neq0$ and $h\neq0$, the model becomes quantumly chaotic \cite{atas-tfim_chaos-prl,jonay-tfim-chaos-arxiv,Roberts-tfim_chaos-jhep}. As usual, the total system $\mathcal{S}$ is initialized in the ground state of the full Hamiltonian Eq.~(\ref{eq:chaotic_tfim}) followed by a sudden quench via the local operator $O = \sigma_{1}^{x}$. We choose subsystem $\mathcal{A}$ to consist of just the first spin and subsystem $\mathcal{B}$ consists of the rest $N-1$ spins. Figure~(\ref{fig:subsystem qfi heatmap}) shows the heatmap for $I_{F}(t,h)$. We observe that $I_{F}(t,h)$ peaks around $h=\pm J$. Note that even though the energy gap never fully closes, it does have a minimum finite energy gap in the presence of a longitudinal field~\cite{sachdev-qpt-book} which according to Eq.~\eqref{eq:scaling_qfi} implies a pseudo-critical amplification of $I_{F}(t,h)$ around $h=\pm J$. Furthermore, $I_{F}(t,h)$ also decays with time, which has been previously observed in Ref.~\cite{tang-time_estimation_scipost} where for subsystem sizes less than half the size of the system, the authors found an exponential decay of $I_{F}(t)$ at intermediate times before it saturates. This behavior reflects the progressive loss of local information due to scrambling. As interactions spread information across the system, the reduced state of a small subsystem approaches a locally equilibrated state, resulting in reduction of distinguishability of nearby states under time evolution and hence the decay of $I_{F}(t)$.

\begin{figure}
\centering
\includegraphics[width=\linewidth]{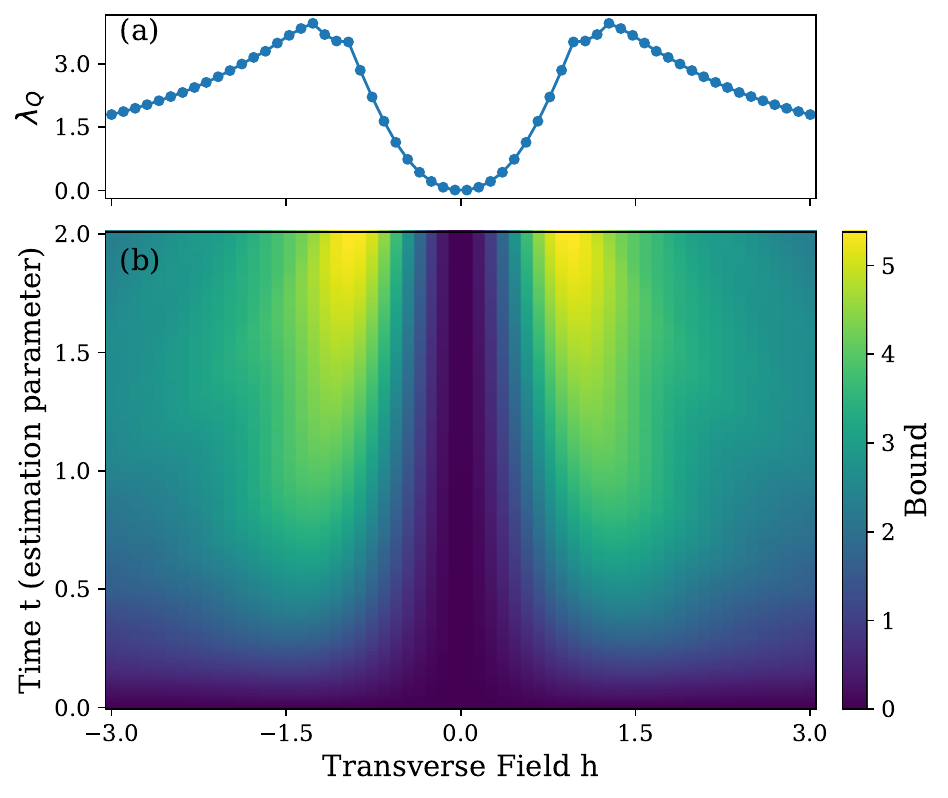}
\caption{(a) Quantum Lyapunov Exponent $\lambda_Q$ as a function of the transverse field $h$. $\lambda_Q$ peaks approximately around $h=\pm J$ demonstrating maximal chaos. (b) Bound on the Lyapunov exponent $\frac{1}{2t}\left(\int_0^t\sqrt{I_F(s,h)}\,ds\right)^2$ as a heatmap. The bound also peaks around the same region as $\lambda_Q$ corresponding to maximal chaos. The results are obtained for the one-dimensional TFIM with a longitudinal field with open boundary conditions. The parameters are $J=1.0$, $g=0.4$, $N=11$.}
\label{fig:fisher-lyapunov heatmap}
\end{figure}

In Fig.~\ref{fig:fisher-lyapunov heatmap}(a), we plot the Lyapunov exponent $\lambda_Q$ extracted from the $\text{OTOC}(t) = \langle \sigma^z_2(t)\sigma^x_1\sigma^z_2(t)\sigma^x_2\rangle$ where the expectation value is taken with respect to the initial ground state. We observe that $\lambda_Q$ exhibits a peak around $h=\pm J$, which is consistent with our bound Eq.~\eqref{eq:final_lyapunov_bound}. 

It is worth emphasizing that in the preceding analysis, we employed the averaged OTOC $\mathcal{O}_t$, as defined in Eq.~\eqref{eq:avg-otoc}, to obtain a universal relation between subsystem QFI and scrambling dynamics which eliminates operator dependent fluctuations. In the numerical simulations, however, we extract $\lambda_Q$ from the conventional OTOC with specific local operators. In chaotic systems, the exponential growth of OTOCs is largely independent of the specific choice of operators \cite{shen-lyapunov_peak-prb} and therefore, the $\lambda_Q$ extracted from $\mathcal{O}_t$ faithfully captures the behavior of quantum chaos across a quantum critical point \cite{shen-lyapunov_peak-prb}. As a final verification of our analytical predictions, we plot our bound Eq.~(\ref{eq:final_lyapunov_bound}) as a heatmap in Fig.~\ref{fig:fisher-lyapunov heatmap}(b) depicting the peak of $\lambda_Q$ being constrained by the $I_{F}(t,h)$.


\paragraph{Concluding remarks}

In the present work, we have developed a metrological perspective on information scrambling by deriving a generalized quantum Cramer-Rao bound. This result elevates the QFI from a standard tool of parameter estimation to a dynamical resource that quantitatively limits scrambling: the more sensitively a subsystem encodes time, the faster the global system can scramble information, as witnessed by the decay of the OTOC. In the case of a single-qubit subsystem, we further derived an inequality that bounds the quantum Lyapunov exponent by the time-integrated subsystem QFI. This relation makes the intuition precise that Lyapunov growth is not an independent dynamical constant, it is limited by the metrological distinguishability of nearby reduced states accessible to a local observer.

A central implication of our framework is a mechanism for the empirically observed enhancement of chaos near quantum criticality. We proved rigorously that the subsystem QFI is amplied near criticality by relating it to an imaginary-time two-point correlation function. In this way, the critical peak of the subsystem QFI provides a direct route to understanding why the quantum Lyapunov exponent can be maximized in the vicinity of a quantum critical point: criticality amplifies local temporal distinguishability, and our bound on the quantum Lyapunov exponent then forces the scrambling rate to reflect this amplification. 

This viewpoint naturally suggests an interpretation in terms of \emph{local quantum stopwatches}. The reduced state $\rho_\mathcal{A}(t)$ encodes time through its distinguishability. The subsystem QFI with time as the parameter, thus, quantifies the ultimate precision with which $A$ can function as a stopwatch. Our results show that the same quantity that makes $A$ a precise stopwatch, i.e, a large $I_F(t)$, also constrains the global scrambling rate. In particular, chaotic dynamics near a quantum critical point make $\rho_A(t)$ exceptionally time-sensitive, so that a small subsystem becomes a naturally emergent most precise stopwatch. 

Reference~\cite{quantum_clock_pra} proposed an interferometric protocol where a controlled ancilla attached to the many-body system of interest acts as a quantum clock. In the Supplemental Material \cite{sm}, we show that in such a setup, our bound becomes an equality. This conceptual picture connects directly to recent experimental realizations of quantum stopwatches based on coherent superpositions of many Rydberg energy eigenstates~\cite{quantum_stopwatch_prr}. Inspired by this approach, one may envision an array of Rydberg atoms engineered to simulate a chaotic many-body Hamiltonian \cite{monterio-chaos_rydberg-prl,turner-scars_rydberg-prx,werman-chaos_rydberg-arxiv}, where local subsystems act as intrinsic quantum stopwatches whose precision is governed by their QFI. Such platforms offer a promising route to experimentally probe the metrology-scrambling connection established in the present work, and to test the predicted enhancement of clock precision and Lyapunov growth near quantum criticality. 

\paragraph{Data Availability} The code for our numerical simulation can be found at \cite{github}

\acknowledgements{F.C. acknowledges funding by the European Union (EQC, 101149233). S.D. acknowledges support from the John Templeton Foundation under Grant No. 63626. This work was supported by the U.S. Department of Energy, Office of Basic Energy Sciences, Quantum Information Science program in Chemical Sciences, Geosciences, and Biosciences, under Award No. DE-SC0025997.}

\bibliography{biblio}

\clearpage
\appendix
\onecolumngrid

\input{supp}
\end{document}

%% file: supp.tex
\clearpage
\onecolumngrid

\section*{Supplemental Material}

\vspace{1em}

\section{Bound for quantum Lyapunov Exponent }

In this section, we derive our bound for the quantum Lyapunov exponent in terms of the time-averaged subsystem QFI $I_{F}(t)$ which is Eq.~(13) of the main text. The starting point is our generalized quantum Cramer-Rao bound 
\begin{equation}\label{eq:gqcr_supp}
    (\Delta\rho_A)^2 \geq \frac{\dot{\mathcal{O}}_t^{2}}{4I_{F}(t)}
\end{equation}
Our quantum many-body system $\mathcal{S}$ is partitioned into two subsystems $\mathcal{A}$ and $\mathcal{B}$. When $\mathcal{A}$ is a single qubit, we can write the reduced density matrix for $\mathcal{A}$ using the Bloch vector representation, as $\rho_{\mathcal{A}}(t) = \frac{1}{2} \left( I + \vec{r}(t) \cdot \vec{\sigma} \right)$. As usual, \( \vec{r}(t) \in \mathbb{R}^3 \) is the Bloch vector with norm \( \abs{\vec{r}(t)} \leq 1 \), and \( \vec{\sigma} = (\sigma_x, \sigma_y, \sigma_z) \) are the Pauli matrices.

The eigenvalues of this qubit state are: $\lambda_{\pm} = (1\pm\abs{\vec{r}})/2$. Thus, we can write $\tr{\rho_\mathcal{A}^2} = (1+\abs{r}^2)/2 = p$ and $\tr{\rho_\mathcal{A}^3} = (3p-1)/2$. Hence, the variance simply reads
\begin{equation}
 (\Delta\rho_{\mathcal{A}})^2 =\frac{1}{2} \left(3\tr{\rho_{\mathcal{A}}^2} - 1\right) - \tr{\rho_{\mathcal{A}}^2}^2
\end{equation}
Now, using the OTOC-RE theorem \cite{fan-otoc_re-scibull} i.e., $\mathcal{O}_t = \exp{-S_{\mathcal{A}}^{(2)}(t)}$, we can rewrite our generalized quantum Cramer-Rao bound Eq.~(\ref{eq:gqcr_supp}) as,
\begin{equation}
\label{eq:qfi-otoc-inequal}
4I_F(t)\ge\frac{2 \dot{\mathcal{O}}_t^{2}}{\left(3\mathcal{O}(t) - 1\right) - 2 \mathcal{O}_t^2}
\end{equation}
This differential inequality can be solved excatly. To this end, we write
\begin{equation}
    \int_{\mathcal{O}_0}^{\mathcal{O}_t}\frac{d\mathcal{O}}{\sqrt{-\mathcal{O}^{2}+\alpha\mathcal{O}+\beta}} \leq 2\int_{0}^{t}\sqrt{I_{F}(s)}ds
\end{equation}
where $\alpha = 3/2$ and $\beta = - 1/2$. Integrating both sides we obtain,
\begin{equation}\label{eq:qsl otoc qudit}
\mathcal{O}_t\leq \frac{1}{4}\cos{\left(2\int_0^t ds\,\sqrt{I_{F}(s)}\right)} + \frac{3}{4}\,. 
\end{equation}
Here we used that $\mathcal{O}_0=1$ and we only consider arguments of the cosine in the fist quadrant, that is early enough times $t$.

For chaotic dynamics, we can further write $\mathcal{O}_t = \exp{-\lambda_Q t}$. Now expanding the cosine up to leading order, we immediately obtain
\begin{equation}\label{eq:lambda-lb-quadratic}
\lambda_Q \ge \frac{1}{2 t}\,\left(\int_0^t\sqrt{I_{F}(s)}ds\right)^2\,.
\end{equation}






\section{Expressing subsystem QFI as imaginary-time correlation functions}

In the previous section, we derived our bound Eq.~(\ref{eq:lambda-lb-quadratic}). This implies that any physical mechanism that amplifies $I_{F}(t)$, in turn, amplifies $\lambda_Q$. Here, we show that $I_{F}(t)$ exhibits critical enhancement in systems that undergo a quantum phase transition. In the following, we express $I_{F}(t)$ in terms of imaginary-time two-point correlation functions, which are known to show universal scaling behavior near quantum critical points. We start by relating $I_{F}(t)$ to the operator $M_{\mathcal{A}}(t) = \ptr{\mathcal{B}}{[H,\rho(t)]}$ using the following relation,
\begin{equation}
\label{eq:qfi-sandwich_rewrite_supp}
    \frac{1}{2p_{\max}}\norm{M_{\mathcal A}(t)}_2^2\ \le\ I_F(t)\ \le\ \frac{1}{2p_{\min}}\norm{M_{\mathcal A}(t)}_2^2,
\end{equation}
The latter follows directly from the definition of the subsystem QFI \eqref{eq:qfi},
\begin{equation}
   I_{F}(t)=2\sum_{jk}^{d}\frac{\abs{\bra{j}\ptr{\mathcal{B}}{[H,\rho(t)]}\ket{k}}^{2}}{p_j+p_k}
\end{equation}
We can write $2p_{\min} \leq p_j+p_k\leq 2p_{\max}$ where $p_{\min(\max)}$ is the minimum (maximum) eigenvalue of $\rho_{\mathcal{A}}(t)$
Assume that the total Hamiltonian of the system $\mathcal{S}$ can be decomposed as $H=H_{\mathcal A}+H_{\mathcal B}+H_{\mathcal I}$, which allows us to write,
\begin{align*}
    M_{\mathcal{A}}(t) &= \ptr{\mathcal{B}}{[H,\rho(t)]} \\
    &=[H_A,\rho_A(t)] + \ptr{\mathcal{B}}{[H_B + H_I,\rho(t)]} \\
    &=\text{unitary term in } \mathcal{A} + \ptr{\mathcal{B}}{[H_{\bar{A}},\rho(t)]} 
\end{align*}
The first term generates a unitary rotation on $\mathcal A$ and does not encode the influence of critical fluctuations in $\mathcal B$ and therefore, we are only interested in the contribution of $H_{\bar{A}}$ to $M_{\mathcal{A}}(t)$. We now wish to prove the following relation
\begin{equation}
\label{eq:qfi-corr_rewrite}
     \norm{M_{\mathcal A}(t)}_2^2 \ \le\
      d_B\!\int_{0}^{\infty}\!\!\! d\tau \int_{0}^{\infty}\!\!\! d\tau'\,
      \mathcal{K}(\tau,\tau')\,
      e^{-it\partial_{\tau}}\,e^{it\partial_{\tau'}}\,
      \tr\!\left\{J(\tau)^{\dagger}J(\tau')\right\},
\end{equation}
where we define the corresponding current operator in imaginary time,
\begin{equation}
\label{eq:img_t_corr}
J(\tau):=[\bar H_{\mathcal A}(\tau),\rho(0)],\qquad
\bar H_{\mathcal A}(\tau):=e^{\tau H}\bar H_{\mathcal A}e^{-\tau H}.
\end{equation} and $\mathcal{K}$ is the inverse Laplace kernel defined by
\begin{equation}
\label{eq:K_def_Laplace_rewrite}
\int_{0}^{\infty}\!\! d\tau \int_{0}^{\infty}\!\! d\tau'\;
\mathcal{K}(\tau,\tau')\, e^{-\Delta\tau}\, e^{-\Delta'\tau'} \;=\; 1.
\end{equation}

First, we will obtain an expression for the LHS of Eq.~(\ref{eq:qfi-corr_rewrite}) and then equate the expression to the RHS. For the rest of this derivation, we will use the eigenbasis of the full Hamiltonian $H$ as our default choice of basis for spectral expansion. Given that the system $\mathcal{S}$ starts in a globally pure state, its time evolved state is given by $\ket{\psi} = \sum_n c_n e^{-iE_n t}\ket{n}$. The corresponding density matrix is $\rho(t) = \sum_{m,n}c_mc_n^*e^{-i\Delta_{mn}t}\ket{m}\bra{n}$. Next we expand our Hamiltonian of interest $H_{\bar{A}} = \sum_{p,q} = (H_{\bar{A}})_{pq}\ket{p}\bra{q}$. We now compute the commutator
\begin{align}
    [H_{\bar A}, \rho(t)]
    &= H_{\bar A} \rho(t) - \rho(t) H_{\bar A}.
\end{align}
For the first time, we obtain
\begin{align}
    H_{\bar A}\rho(t)
    &= \left( \sum_{p,q} (H_{\bar A})_{pq} \ket{p}\bra{q} \right)
       \left( \sum_{m,n} c_m c_n^* e^{-i \Delta_{mn} t} \ket{m}\bra{n} \right) \\
   &= \sum_{p,q} \sum_{m,n} (H_{\bar A})_{pq}
       c_m c_n^* e^{-i \Delta_{mn} t} \ket{p} \underbrace{\bra{q}\ket{m}}_{\delta_{qm}} \bra{n} \\
    &= \sum_{m,n} \sum_{p} (H_{\bar A})_{pm}
       c_m c_n^* e^{-i \Delta_{mn} t} \ket{p}\bra{n}.
\end{align}
The second term can be manipulated to read,
\begin{align}
    \rho(t) H_{\bar A}
    &= \left( \sum_{m,n} c_m c_n^* e^{-i \Delta_{mn} t} \ket{m}\bra{n} \right)
       \left( \sum_{p,q} (H_{\bar A})_{pq} \ket{p}\bra{q} \right) \\
    &= \sum_{m,n} \sum_{p,q} c_m c_n^* e^{-i \Delta_{mn} t}
       (H_{\bar A})_{pq} \ket{m} \underbrace{\bra{n}\ket{p}}_{\delta_{np}} \bra{q} \\
    &= \sum_{m,n} \sum_{q} c_m c_n^* e^{-i \Delta_{mn} t}
       (H_{\bar A})_{nq} \ket{m}\bra{q}.
\end{align}
Collecting all terms the commutator becomes
\begin{align}
    [H_{\bar A}, \rho(t)]
    &= H_{\bar A}\rho(t) - \rho(t) H_{\bar A} \\
    &= \sum_{m,n} c_m c_n^* e^{-i\Delta_{mn}t} 
       \Big[ \sum_p(H_{\bar A})_{pm} \ket{p}\bra{n}
           -  \sum_q(H_{\bar A})_{nq}\ket{m}\bra{q} \Big],
\end{align}
and therefore
\begin{align}\label{eq:MA_inequal}
    \norm{M_{\mathcal{A}}(t)}_2^2 &= \tr_A\Big(\tr_{\mathcal{B}}[H_{\bar A}, \rho(t)]^{\dagger}\tr_{\mathcal{B}}[H_{\bar A}, \rho(t)]\Big) \leq d_{\mathcal{B}}\tr_{{\mathcal{AB}}}\Big([H_{\bar A}, \rho(t)]^{\dagger}[H_{\bar A}, \rho(t)]\Big)\\
\end{align}
where $d_{\mathcal{B}} = \mathrm{dim}\,\mathcal{H_{\mathcal{B}}}$ is the dimension of the Hilbert space of subsystem $\mathcal{B}$.
Plugging in the expressions for the commutator derived above, inspect the first term on the RHS of Eq.~(\ref{eq:MA_inequal}), namley we have
\begin{align}\label{eq:1term}
    &=d_B\Bigg(\sum_{m,n,m',p'} c_m^*c_nc_m'c_n^*e^{-it(\Delta_{m'n}-\Delta_{mn})}(H_{\bar A})_{pm}(H_{\bar A})_{pm'}\Bigg)
\end{align}
Next, we will show that we get the same term with kernel transformations using the RHS of Eq.~(\ref{eq:qfi-corr_rewrite}). Using spectral expansion, we define
\begin{equation}
    J(\tau+it) = \big[H_{\bar A}(\tau+it), \rho(0)\big],
\end{equation}
with
\begin{equation}
    H_{\bar A}(\tau+it) = e^{(\tau+it)H} \, H_{\bar A} \, e^{-(\tau+it)H}.
\end{equation}
to obtain
\begin{align}
    H_{\bar A}(\tau+it)
    &= e^{(\tau+it)H} \left( \sum_{p,q} (H_{\bar A})_{pq} \ket{p}\bra{q} \right)
       e^{-(\tau+it)H} \\
    &= \sum_{p,q} (H_{\bar A})_{pq} \, e^{(\tau+it)E_p} \ket{p}\bra{q} e^{-(\tau+it)E_q} \\
    &= \sum_{p,q} (H_{\bar A})_{pq} \,
       e^{(\tau+it)(E_p - E_q)} \ket{p}\bra{q} \\
    &= \sum_{p,q} (H_{\bar A})_{pq} \,
       e^{(\tau+it)\Delta_{pq}} \ket{p}\bra{q}.
\end{align}

We now compute the commutator
\begin{equation}
    J(\tau+it) = H_{\bar A}(\tau+it)\,\rho(0) - \rho(0)\,H_{\bar A}(\tau+it).
\end{equation}
As before, the first term, $H_{\bar A}(\tau+it)\rho(0)$, becomes 
\begin{align}
    H_{\bar A}(\tau+it)\rho(0)
    &= \left( \sum_{p,q} (H_{\bar A})_{pq} e^{(\tau+it)\Delta_{pq}} \ket{p}\bra{q} \right)
       \left( \sum_{m,n} c_m c_n^* \ket{m}\bra{n} \right) \\
    &= \sum_{p,q} \sum_{m,n}
       (H_{\bar A})_{pq} e^{(\tau+it)\Delta_{pq}}
       c_m c_n^* \ket{p} \underbrace{\bra{q}\ket{m}}_{\delta_{qm}} \bra{n} \\
    &= \sum_{p,m,n} (H_{\bar A})_{pm} e^{(\tau+it)\Delta_{pm}}
       c_m c_n^* \ket{p}\bra{n}.
\end{align}
The second term, $\rho(0)H_{\bar A}(\tau+it)$, can be written as
\begin{align}
    \rho(0)H_{\bar A}(\tau+it)
    &= \left( \sum_{m,n} c_m c_n^* \ket{m}\bra{n} \right)
       \left( \sum_{p,q} (H_{\bar A})_{pq} e^{(\tau+it)\Delta_{pq}} \ket{p}\bra{q} \right) \\
    &= \sum_{m,n} \sum_{p,q}
       c_m c_n^* (H_{\bar A})_{pq} e^{(\tau+it)\Delta_{pq}}
       \ket{m}\bra{n}\ket{p}\bra{q} \\
    &= \sum_{m,n} \sum_{p,q}
       c_m c_n^* (H_{\bar A})_{pq} e^{(\tau+it)\Delta_{pq}}
       \ket{m} \underbrace{\braket{n|p}}_{\delta_{np}} \bra{q} \\
    &= \sum_{m,n,q}
       c_m c_n^* (H_{\bar A})_{nq} e^{(\tau+it)\Delta_{nq}}
       \ket{m}\bra{q}.
\end{align}
This, finally the flux $J(\tau+it)$ can be obtained.Combining the two contributions, we have
\begin{align}
    J(\tau+it)
    &= H_{\bar A}(\tau+it)\rho(0) - \rho(0)H_{\bar A}(\tau+it) \\
    &= \sum_{m,n}c_mc_n^*
       \Big(\sum_p (H_{\bar A})_{pm} e^{(\tau+it)\Delta_{pm}}
        \ket{p}\bra{n}
    -\sum_{q}
        (H_{\bar A})_{nq} e^{(\tau+it)\Delta_{nq}}
       \ket{m}\bra{q}\Big).
\end{align}
Plugging this expansion into the RHS of Eq (\ref{eq:qfi-corr_rewrite}), we obtain four terms. Now, inspecting the first term,
\begin{align}
    &=\sum_{m,n,m',p} c_m^*c_nc_{m'}c_{n'}^{*} H_{pm}H_{pm'}e^{-it(\Delta_{pm}-\Delta_{pm'})}e^{(\Delta_{pm}+\Delta_{pm'})\tau}\\
    &=\sum_{m,n,m',p} c_m^*c_nc_{m'}c_{n'}^{*} H_{pm}H_{pm'}e^{-it(\Delta_{m'n}-\Delta_{mn})}e^{(\Delta_{pm}+\Delta_{pm'})\tau}
\end{align}
We immediately observe that this is the same as Eq.~(\ref{eq:1term}), which is the 1st term in the expansion of LHS of Eq.~(\ref{eq:qfi-corr_rewrite}), except the extra Laplace-like Kernel term. The same can be shown for the rest of the three terms. This implies that if we multiply $\tr{J(\tau+it)^{\dagger}J(\tau'+it)}$ with some inverse kernel $\mathcal{K}(\tau,\tau')$, then it becomes equal to RHS of Eq.~(\ref{eq:qfi-corr_rewrite}). In conclusion, we have,
\begin{align}\label{eq:qfi-imag-time}
    \norm{M_{\mathcal{A}}(t)}^{2}_{2}&\leq \int\int d\tau d\tau'\mathcal{K}(\tau,\tau')\tr{J(\tau+it)^{\dagger}J(\tau+it)}\\ \nonumber
    &= \int\int d\tau d\tau'\mathcal{K}(\tau,\tau')e^{-it\frac{\partial}{\partial\tau}}e^{it\frac{\partial}{\partial\tau'}}\tr{J(\tau)^{\dagger}J(\tau')}
\end{align}

\section{Scaling Analysis of subsystem QFI}

Having expressed $M_{\mathcal{A}}(t)$ in terms of imaginary-time two-point correlators, we now show that these correlation functions exhibit universal scaling near a quantum critical point. Note that the only step producing an inequality in Eq.~\eqref{eq:qfi-corr_rewrite} is the Hilbert-Schmidt bound under partial trace,
$\|\Tr_{\mathcal B}X\|_2^2\le d_B\|X\|_2^2$ which contributes non-universal prefactors.

We start by defining the intensive quantity
\begin{equation}
m_A^2(t) := \frac{\|M_A(t)\|_2^2}{L^d}
\label{eq:intensive}
\end{equation}
Now assume that $ H_{\bar{A}}$ is a sum of local densities $ H_{\bar{A}} = \sum_x h(x)$. Then the current operators can also be decomposed in terms of local current densities
\begin{equation}
J(\tau)
=
\sum_x j(x,\tau),
\qquad
j(x,\tau) := [h(x,\tau), \rho(0)]
\end{equation}
Therefore,
\begin{equation}
\mathrm{tr}\!\left[J^\dagger(\tau) J(\tau')\right]
=
\sum_{x,y}
\langle
j^\dagger(x,\tau) j(y,\tau')
\rangle.
\end{equation}
Now we assume translation invariance, so that $\langle j^\dagger(x,\tau) j(y,\tau') \rangle
= C_{jj}(x-y;\tau,\tau')$ which allows us to write
\begin{align}
\sum_{x,y} C_{jj}(x-y;\tau,\tau')
&=
\sum_y \sum_r C_{jj}(r;\tau,\tau')
\\
&=
L^d \sum_r C_{jj}(r;\tau,\tau')
\end{align}
Thus $\tr{J^\dagger(\tau) J(\tau')}\sim L^d \sum_r C_{jj}(r;\tau,\tau')$.

Dividing by $L^d$ and going to the continuous space $\sum_r \rightarrow \int d^d x$
we obtain
\begin{equation}
m_A^2(t)
\sim
\int d^d x
\int d\tau d\tau'
K(\tau,\tau')
e^{-it\partial_\tau}
e^{it\partial_{\tau'}}
C_{jj}(x;\tau,\tau').
\label{eq:scaling_structure}
\end{equation}
Under dynamical scaling $x \to \alpha x$ and $\tau \to \alpha^\zeta \tau$, we have $\partial_\tau \to \alpha^{-\zeta} \partial_\tau$. If $t \to \alpha^\zeta t$, then $t \partial_\tau \to t \partial_\tau$, so the shift operators are RG-invariant. Assume the kernel $K(\tau,\tau')$ is non-singular at criticality. Then the leading scaling is governed solely by
\begin{equation}
m_A^2(t)
\sim
\int d^d x \int d\tau d\tau'
C_{jj}(x;\tau,\tau').
\label{eq:core_scaling}
\end{equation}

Now, assume $j(x,\tau)$ has scaling dimension $\Delta_j$,
\begin{equation}
j(x,\tau)
\to
\alpha^{-\Delta_j}
j\!\left(\frac{x}{\alpha}, \frac{\tau}{\alpha^\zeta}\right).
\end{equation}
Then
\begin{equation}
C_{jj}(x;\tau,\tau')
\to
\alpha^{-2\Delta_j}
C_{jj}\!\left(\frac{x}{\alpha};
\frac{\tau}{\alpha^\zeta},
\frac{\tau'}{\alpha^\zeta}\right).
\end{equation}
Under the change of variables $x = \alpha x',
\quad
\tau = \alpha^\zeta \tau_1,
\quad
\tau' = \alpha^\zeta \tau_2 $
the measure scales as $d^d x\, d\tau\, d\tau'
=\alpha^{d+2\zeta}d^d x' d\tau_1 d\tau_2$.
This gives us,
\begin{equation}
m_A^2(t)
\to
\alpha^{d+2\zeta}
\alpha^{-2\Delta_j}
m_A^2(t)
=
\alpha^{d+2\zeta - 2\Delta_j}
m_A^2(t).
\end{equation}

Define $\Delta_M$ via $m_A^2 \to \alpha^{-\Delta_M} m_A^2$, where $\Delta_M=2\Delta_j - d - 2\zeta$. Let $\xi$ be the correlation length, then one has $\xi = \abs{\lambda-\lambda_c}^{-\nu}$, where $\nu$ is the correlation length critical exponent and let $\Delta_{\lambda}$ be the scaling dimensions of the tuning parameter $\lambda$, $\nu = \Delta_{\lambda}^{-1}$. Let $\delta\lambda = \lambda - \lambda_c$ with scaling dimension $\Delta_\lambda$: $\delta\lambda \to \alpha^{\Delta_\lambda} \delta\lambda$. Choose $\alpha$ such that $1 = \alpha^{\Delta_\lambda} |\delta\lambda|,\qquad\alpha =|\delta\lambda|^{-1/\Delta_\lambda}$. Then $m_A^2(\delta\lambda)\sim\alpha^{-\Delta_M}=|\delta\lambda|^{\Delta_M/\Delta_\lambda}$. Hence,
\begin{equation}
m_A^2(t)
\sim
|\lambda - \lambda_c|^{\Delta_M/\Delta_\lambda},
\qquad
\Delta_M = 2\Delta_j - d - 2\zeta.
\label{eq:26_final}
\end{equation}

The subsystem time-QFI inherits the same critical exponent via the following equation
\begin{equation}\label{eq:final_lyapunov_bound_final_supp}
I_F^A(t)
\sim
|\lambda - \lambda_c|^{\Delta_M/\Delta_\lambda}.
\end{equation}
Equations~(\ref{eq:lambda-lb-quadratic}) and (\ref{eq:final_lyapunov_bound_final_supp}) imply that the quantum Lyapunov exponent achieves a critical amplification across the quantum critical point whenever $\Delta_M<0$. This leads to the physical picture that quantum chaotic systems undergoing a quantum phase transition naturally act as the most precise quantum stopwatch.

\section{Quantum Clocks}

Finally, we outline a scenario, for which our bound on the quantum Lyapunov exponent () becomes tight. In Ref.~\cite{quantum_clock_pra}, the authors build a quantum clock by coupling an ancilla qubit globally to a many body system of interest. The ancilla qubit then acts as a quantum clock in the sense that whenever it is in state $\ket{0}$, the many-body system moves forward in time and whenever the the ancilla is in state $\ket{1}$, the many-body system's Hamiltonian changes sign and hence moves backward in time.

A canonical choice is
\begin{equation}
    H_{\text{tot}}
    \;=\;
    \ket{0}\!\bra{0}_\tau \otimes H
    \;+\;
    \ket{1}\!\bra{1}_\tau \otimes (-H)
    \;=\;
    \tau^z \otimes H \,,
    \label{eq:H_tot}
\end{equation}
where $\tau^z = \ket{0}\bra{0} - \ket{1}\bra{1}$ acts on the ancilla. Thus,
if the ancilla is in $\ket{0}$, the chain evolves forward under $H$, while
if it is in $\ket{1}$, the chain evolves backward under $-H$.

We assume an initial product state
\begin{equation}
    \ket{\Psi_0}
    \;=\;
    \ket{+}_\tau \otimes \ket{\psi}_B \,,
    \qquad
    \ket{+}_\tau = \frac{\ket{0} + \ket{1}}{\sqrt{2}} \,,
\end{equation}
where $\ket{\psi}_B$ is an arbitrary (possibly highly entangled) state of
the chain.

Following the interferometric protocol, we consider two local operators
$O_1$ and $O_2$ acting on the chain, typically supported on distinct spatial
regions. The protocol implements the following sequence on the joint system:
\begin{enumerate}
    \item Prepare $\ket{\Psi_0}$.
    \item Conditioned on the ancilla state, evolve the chain forward and
    backward for a duration $t$ under $H_{\text{tot}}$.
    \item Apply $O_1$ and $O_2$ in a prescribed forward/backward order,
    again controlled by the ancilla.
\end{enumerate}
The resulting final state can be written as
\begin{equation}
    \ket{\Psi_f(t)}
    =
    \frac{1}{\sqrt{2}}\Big(
        \ket{L(t)}_B \otimes \ket{0}_\tau
        +
        \ket{R(t)}_B \otimes \ket{1}_\tau
    \Big),
    \label{eq:Psi_final}
\end{equation}
with branch states
$\ket{R(t)}_B
    =
    e^{+i H t}\, O_2\, e^{-i H t}\, O_1\,\ket{\psi}_B$
    and $\ket{L(t)}_B
    =
    O_1\, e^{+i H t}\, O_2\, e^{-i H t}\,\ket{\psi}_B$.

Tracing out the chain, the reduced density matrix of the ancilla is
\begin{equation}
\label{eq:ancilla_clock_state}
    \rho_A(t)
    \;=\;
    \ptr{B}{\ket{\Psi_f(t)}\bra{\Psi_f(t)}}
    \;=\;
    \frac{1}{2}
    \begin{pmatrix}
        1 & \mathcal{O}_t \\
        \mathcal{O}^*_t & 1
    \end{pmatrix},
\end{equation}
where the off-diagonal coherence is given by
\begin{equation}
    \mathcal{O}_t
    \;=\;
    \bra{L(t)} R(t)\rangle
    \;=\;
    \big\langle
        O_2(t)\, O_1(0)\, O_2(t)\, O_1(0)
    \big\rangle_{\ket{\psi}} \,,
    \label{eq:OTOC_def}
\end{equation}
which is precisely an OTOC in the chain.

Using Eq.~(\ref{eq:ancilla_clock_state}), we can compute the QFI of the ancilla and express it in terms of the OTOC $\mathcal{O}_t$ of the many-body system.,
\begin{equation}
    I_{F} = \frac{ \dot{\mathcal{O}}_t^2}{1 - \mathcal{O}^2_t} 
    \label{eq:F_diff_ineq}
\end{equation}
This  can be solved as before to obtain,
\begin{equation}
    \mathcal{O}_t = \cos\!\left( \int_0^t \! ds\, \sqrt{I_F(s)} \right)
    \label{eq:F_cos_bound}
\end{equation}
For chaotic dynamics $\mathcal{O}_t = \exp{-\lambda_Q t}$ and Taylor expanding to second order in time, we have,
\begin{equation} \label{eq:lambda_QFI_bound_general}
   \lambda_Q = \frac{1}{2t}\left( \int_0^t \sqrt{I_F(s)}\ ds  \right)^2 + \mathcal{O}(t^2).
\end{equation}
This is exactly Eq.~(\ref{eq:final_lyapunov_bound}) we obtained before but with an equality instead of an inequality. In the previous case, the first qubit of the many-body chain was subsystem $\mathcal{A}$ and $\mathcal{B}$ was the rest of the chain. Hence, in some sense the fist spin acted as an effective ancilla qubit for the rest of the chain and hence an effective quantum stopwatch, whose conditional evolution is more complex than the ancilla based quantum clock presented in Ref.~\cite{quantum_clock_pra}.